\advance\voffset by -25mm
\advance\hoffset by -25mm

\documentclass{article}[12pt]
\usepackage[]{graphicx}
\def\betabfeq{{\mbox{\boldmath $ \beta$}}}

\newcounter{saveeqn}

\begin{document}

\title{{ \bf  Reply to "A first principles derivation of the electromagnetic fields of a point charge in arbitrary motion"} }

\author{Young-Sea Huang\\ Department of Physics, Soochow University, Shih-Lin, Taipei 111, Taiwan\\yshuang@mail.scu.edu.tw }

\maketitle

\vspace{25mm}

%\begin{abstract}

%\end{abstract}

\vspace{4cm}



\vfill\eject
\setlength{\baselineskip}{18pt}

  In a recently-published article,  Singal made a hasty conclusion, "Huang and  Lu\cite{Huang1} considered the more general case but found incorrect expressions for the fields".\cite{Singal}  As an author of the article being criticized,  I am obliged to respond  to such conclusion.       
  
According to the current formulation of  the electromagnetic radiation from an accelerated point charge\cite{1,2,3,4}, the electric and magnetic fields of the waves radiated from the point charge are  (in SI units)
\begin{equation}\label{eq35}
{\bf E} ({\bf r},t) 
= { q  \over 4 \, \pi \,\epsilon_{0}\, c }  { \hat{ {\bf r}} \times ( (\hat{ {\bf r} }-  {\betabfeq} ) \times \dot{ \betabfeq} \,  ) \over  r \, (1- {\betabfeq} \cdot  \hat{ {\bf r}})^{3} } , 
 \end{equation}
\begin{equation}\label{eq35b}
{\bf B} ({\bf r},t) 
= {\mu_{0} \, q \,\, \over 4 \, \pi   }   {\hat{ {\bf r}} \times  (\hat{ {\bf r}} \times ( (\hat{ {\bf r} }-  {\betabfeq} ) \times {\dot{ \betabfeq}} \,  )\, ) \over r\,   (1- {\betabfeq} \cdot  \hat{ {\bf r}})^{3} } . 
 \end{equation}
Here,  $\epsilon_{0}$ and  $\mu_{0}$ are respectively electric permittivity and magnetic permeability of free space, ${\betabfeq}={\bf v}/c$, ${\bf v}$ the velocity of the point
 charge, $q$ the charge,  $r$  the radial distance that the radiated waves travel, and $ \hat{ {\bf r}}$  a unit vector indicating the direction into which  the infinitesimal portion of the waves radiate. 
The electric and magnetic fields of the waves at a position $ {\bf r}$ and  time $t$ are due to the radiation emanating from the charge at the earlier time $t_{r}=t-r/c$.  Thus,  the quantities  $\betabfeq$ and $\dot{ \betabfeq}$ in the right-hand  side of Eqs.~(\ref{eq35}) and (\ref{eq35b}) are to be evaluated  at the retarded time $t_{r}=t-r/c$.  Without using the Li\'{e}nard-Wiechert potential,  rather by the relativistic transformations of velocity, acceleration and electromagnetic fields, Singal obtained the same  acceleration field of radiation;  the acceleration field of  Eq.~(37) in the reference \cite{Singal} is the same as Eq.~(\ref{eq35}).

 However,  starting with  the electromagnetic fields of radiation from  a point charge being instantaneously at rest but accelerated,  and  using  the same relativistic transformations,  we obtained the electric and magnetic fields of the waves radiated from a point charge in arbitrary motion \cite{Huang1, Huang2}
\begin{equation}\label{huangeq35}
{\bf E} ({\bf r},t) 
=  { q  \over 4 \, \pi \,\epsilon_{0}\, c }   {\gamma \,  \hat{ {\bf r}} \times ( (\hat{ {\bf r} }-  {\betabfeq} ) \times \dot{ \betabfeq} \,  ) \over  r \, (1- {\betabfeq} \cdot  \hat{ {\bf r}})^{2} } , 
  \end{equation}
\begin{equation}\label{huangeq35b}
{\bf B} ({\bf r},t) 
= {\mu_{0} \, q \,\, \over 4 \, \pi   }   {\gamma \,  \hat{ {\bf r}} \times  (\hat{ {\bf r}} \times ( (\hat{ {\bf r} }-  {\betabfeq} ) \times {\dot{ \betabfeq}} \,  )\, ) \over r\,   (1- {\betabfeq} \cdot  \hat{ {\bf r}})^{2} } , 
 \end{equation}
 where $\gamma = 1/\sqrt{1- \beta ^2}$. 
One  remarkable distinction between these electromagnetic fields of radiation of  the two different derivations  is the factor $\gamma$.  It seems very dubious that  Eqs.~(\ref{eq35}) and (\ref{eq35b})  lack  the $\gamma$ factor,  since  all the  relativistic transformations of  velocity,  acceleration and electromagnetic fields contain the $\gamma$ factor. 

  We will argue that the currently-accepted expressions Eq.~(\ref{eq35}) and Eq.~(\ref{eq35b}) are probably incorrect.
From these equations, the energy flux density of radiation is 
\begin{equation}\label{eq36}
 {\bf S} ({\bf r},t) 
= { q^{2}  \over 16 \, \pi^{2}\, \epsilon_{0}  \, c}   { (  \hat{ {\bf r}} \times (\, (\hat{ {\bf r} }-  {\betabfeq} ) \times \dot{\betabfeq}\,  )\, )^{2} \over  r^{2} \, (1- {\betabfeq} \cdot  \hat{ {\bf r}} )^{6} }  \,\, \hat{ {\bf r}}. 
\end{equation}
 The energy  radiated into a solid angle $d \Omega$ in the direction $\hat{ {\bf r}}$, and then measured at the position ${\bf r}$ and the time $t$ is $dW({\bf r},t) ={\bf S} \cdot {\hat {\bf r}} \,  r^{2}\, d \Omega \,dt$,  for an infinitesimal time interval $dt$. Here ${\bf S} \cdot {\hat {\bf r}} $ is the energy per unit area per unit time detected at the  position $ {\bf r}$ and the time $t$ of the radiation emitted from  the position of the charge at the retarded time $t_{r}=t-r/c$ .  Hence,  the power radiated per unit solid angle passing through  a surrounding sphere of radius $r$ at the time $t$ is
\begin{equation}\label{eqjak3}
 {d\, P({\hat {\bf r}},t) \over d\, \Omega}= \frac{ d\, W ({\bf r},t) }{ d\, \Omega \,\, d\, t } = {q^{2} \over 16 \, \pi^{2} \,\epsilon_{0} \, c}   { ( \, \hat{ {\bf r}} \times (\, (\hat{ {\bf r} }-  {\betabfeq} ) \times \dot{\betabfeq} \,  )\, )^{2} \over  (1- {\betabfeq} \cdot  \hat{ {\bf r}} )^{6} }. 
 \end{equation}
Eq.~(\ref{eqjak3}) is the power radiated per unit solid angle measured by observers on the  surrounding sphere of radius $r$.\cite{4}
By integrating Eq.~(\ref{eqjak3}) over the  surrounding sphere, the total radiated power which passes through the sphere at the time $t$ is 
\begin{equation}\label{eq38b}
P(t) = {q^{2} \, a^{2} \, \gamma ^{8} \over 6 \, \pi \,\epsilon_{0}\, c^3}\, ( {5+\beta^{2} \over 5} - {4\beta^{2} + 2\beta^{4} \over 5} \, sin^{2} \alpha \,) , 
\end{equation}
where $\alpha$ is the angle between the velocity and the acceleration ${\bf a}$ of the charge at the retarded time $t_{r}$.
Yet,  the total  power radiated by the charge at the retarded time $t_{r}$ is given by the  Li\'{e}nard formula,\cite{1,2,3,4}
\begin{equation}\label{eq38s}
P(t_{r}) = { q^{2} \, \gamma^{6}  \over 6 \, \pi \,\epsilon_{0}\, c^3}   (  {\bf a}^{2} - ( {\betabfeq} \times  {\bf a} )^{2} \,) = { q^{2} \, a^{2} \, \gamma ^{6} \over 6 \, \pi \,\epsilon_{0}\, c^3}\, ( 1- \beta^{2} \, sin^{2} \alpha \,) . 
\end{equation}

The total radiated power that passes through any surrounding sphere is $P(t)$.  Yet, $P(t)$  is different from the total  power  radiated by the charge at the retarded time $P(t_{r})$.  This violates the conservation of energy; the reason is as follows: 
 The total radiated power that passes through a sphere of smaller radius $r'\, (r'<r)$ at an earlier time $t'\, (t'<t)$ is equal to the total radiated power that passes through a sphere of radius $r$ at the time $t$, since the radiation emitted by the charge expands spherically. As the sphere  is getting smaller, and reduced to the position of the charge at the retarded time $t_{r}$, $P(t)$ becomes the total power  instantaneously radiated by the charge at the retarded time $t_{r}$.  Therefore, $P(t)$ must  be equal to  $P(t_{r})$  by the conservation of energy.  This indicates that the electric and magnetic fields of the waves radiated from an accelerated point  charge Eqs.~(\ref{eq35}) and (\ref{eq35b})  are incorrect.
 
 In contrast,  from Eqs.~(\ref{huangeq35}) and~(\ref{huangeq35b}), the energy flux density of electromagnetic radiation is 
\begin{equation}\label{huangeq36}
 {\bf S} ({\bf r},t) 
= { q^{2}  \over 16 \, \pi^{2}\, \epsilon_{0}  \, c}    {\gamma^{2} \, (  \hat{ {\bf r}} \times (\, (\hat{ {\bf r} }-  {\betabfeq} ) \times \dot{\betabfeq} \,  )\, )^{2} \over  r^{2} \, (1- {\betabfeq} \cdot  \hat{ {\bf r}} )^{4} }  \,\, \hat{ {\bf r}}. 
\end{equation}
The power radiated per unit solid angle passing through any surrounding sphere of radius $r(t)$ at any time $t$ is
\begin{equation}\label{huangeqjak3}
 {d\, P({\hat {\bf r}},t) \over d\, \Omega}= { q^{2}  \over 16 \, \pi^{2}\, \epsilon_{0}  \, c}    {\gamma^{2} \, ( \, \hat{ {\bf r}} \times (\, (\hat{ {\bf r} }-  {\betabfeq} ) \times  \dot{\betabfeq} \,  )\, )^{2} \over  (1- {\betabfeq} \cdot  \hat{ {\bf r}} )^{4} }. 
\end{equation}
Then, by  integrating  Eq.~(\ref{huangeqjak3})  over a surrounding sphere,  the total  radiated power passing through  any surrounding sphere is
\begin{equation}\label{huangeq38}
P  = { q^{2} \, \gamma^{6}  \over 6 \, \pi \,\epsilon_{0}\, c^3}   (  {\bf a}^{2} - ( {\betabfeq} \times  {\bf a} )^{2} \,) . 
\end{equation}
The total  radiated power passing through  any surrounding sphere Eq.~(\ref{huangeq38}) is equal to the total power radiated by the charge at the retarded time $P(t_{r})$;  that is,  the principle of conservation of energy is satisfied.  Therefore,   Eqs.~(\ref{huangeq35}) and~(\ref{huangeq35b}) are correct, whereas  Eqs.~(\ref{eq35}) and~(\ref{eq35b}) are not.

 We wish that this reply will initiate further examination on  the currently-accepted formulation of electromagnetic radiation from an accelerated point charge.  Further examination will at least  clarify some ambiguities and misunderstanding in the current formulation,  if the current formulation is indeed correct.

\end{document}